\documentclass[9pt,twocolumn,superscriptaddress,floatfix,nofootinbib]{revtex4-2}


\usepackage{graphicx}%
\usepackage{multirow}%
\usepackage{lipsum}
\usepackage{amsmath}
\usepackage{amsfonts, amssymb, bbm, braket, accents}
\usepackage{graphicx}
\usepackage[usenames,dvipsnames]{color}
\usepackage{soul}
\usepackage{float}
\makeatletter
\let\newfloat\newfloat@ltx
\makeatother
\usepackage{algorithm}
\usepackage{array}
\usepackage[makeroom]{cancel}
\usepackage{booktabs}
\usepackage{xcolor}
\usepackage{siunitx}
\usepackage{float}

\definecolor{cellmin}{rgb}{1,1,1}
\definecolor{cellmax}{rgb}{0.25,1,0.25}

\newcommand{\probP}{\text{I\kern-0.15em P}}
\newcommand{\probE}{\text{I\kern-0.15em E}}

\usepackage[breaklinks=true]{hyperref}
\hypersetup{
  colorlinks   = true,
  urlcolor     = blue,
  linkcolor    = blue,
  citecolor   = red
}
\usepackage{amsthm}

\usepackage[capitalise]{cleveref}
\crefformat{equation}{Eq.~(#2#1#3)}
\crefformat{section}{Sec.~#2#1#3}
\Crefformat{equation}{Equation~(#2#1#3)}
\crefformat{figure}{Fig.~#2#1#3}
\crefrangeformat{equation}{Eqs.~#3(#1)#4--#5(#2)#6}
\Crefformat{section}{Section~#2#1#3}

\begin{document}
\title{Quantum enhanced stratification of Breast Cancer: exploring quantum expressivity for real omics data}

%\author{Lirandë Pira}
%%\email{lirande.pira@student.uts.edu.au}

\author{Valeria Repetto}
\email{valeria.repetto@iit.cnr.it}
\affiliation{National Research Council of Italy (CNR), Institute of Informatics and Telematics (IIT),via Giuseppe Moruzzi 1,Pisa, Italy}
\affiliation{Computational Biomedicine Unit, Department of Medical Sciences, University of Turin,via Santena 19,Torino,Italy}
\author{Elia Giuseppe Ceroni}
%\email{elia.ceroni@student.unisi.it}
\affiliation{National Research Council of Italy (CNR), Institute of Informatics and Telematics (IIT),via Giuseppe Moruzzi 1,Pisa, Italy}
\affiliation{University of Siena,via Roma 56, Siena,Italy}
\author{Giuseppe Buonaiuto}
\affiliation{Institute for High Performance Computing and Networking (ICAR), National Research Council of Italy (CNR), via Pietro Castellino, Naples, Italy}
\author{Romina D'Aurizio}
%\email{romina.daurizio@iit.cnr.it}
\affiliation{National Research Council of Italy (CNR), Institute of Informatics and Telematics (IIT),via Giuseppe Moruzzi 1,Pisa, Italy}

\date{\today}

\begin{abstract}
Quantum Machine Learning (QML) is considered one of the most promising applications of Quantum Computing in the Noisy Intermediate Scale Quantum (NISQ) era for the impact it is thought to have in the near future. Although promising theoretical assumptions, the exploration of how QML could foster new discoveries in Medicine and Biology fields is still in its infancy with few examples. In this study, we aimed to assess whether Quantum Kernels (QK) could effectively classify subtypes of Breast Cancer (BC) patients on the basis of molecular characteristics. We performed an heuristic exploration of encoding configurations with different entanglement levels to determine a trade-off between kernel expressivity and performances. Our results show that QKs yield comparable clustering results with classical methods while using fewer data points, and are able to fit the data with a higher number of clusters. Additionally, we conducted the experiments on the Quantum Processing Unit (QPU) to evaluate the effect of noise on the outcome. We found that less expressive encodings showed a higher resilience to noise, indicating that the computational pipeline can be reliably implemented on the NISQ devices. Our findings suggest that QK methods show promises for application in Precision Oncology, especially in scenarios where the dataset is limited in size and a granular non-trivial stratification of complex molecular data cannot be achieved classically.
\end{abstract}

\maketitle

\section{Introduction}\label{introduction}
Quantum Machine Learning (QML) has been proposed as one of the most promising and worth investigating applications of Quantum Computing (QC) for current NISQ devices\citep{Biamonte2016QuantumML,Dunjko2018MachineL,PerdomoOrtiz2017OpportunitiesAC,Schuld2021QuantumML}. In the QML framework, classical ML algorithms are embedded into the quantum-mechanical formalism by exploiting the paradigm shift of QC's information. The quantum processing leverages entanglement, superposition and interference to enhance, speed up and optimise over classical computing, sometimes even exponentially faster \citep{Bernstein1993QuantumCT,Shor1994AlgorithmsFQ}. This can help in learning in both computational and sample complexity aspects \citep{Huang2020PowerOD}. 
Nevertheless, the quantum embedding alone does not transversally guarantee learnability  \citep{Park2020PracticalAI,GilFuster2023OnTE,Slattery2022NumericalEA} in particular for real classical data \citep{Cerezo2022ChallengesAO}.
Classical ML methods thrive to find and learn patterns in the data and they were successfully applied to different fields, including biology and medicine. Specifically for precision oncology, ML methods are crucial in specific tasks, such as  prognostic predictions, human cancer stratification, and discovery of new clinical subgroups using omics datasets\citep{Tran2021DeepLI}. Nevertheless, despite the substantial progress facilitated by computational methods, some fundamental constraints in modelling biological and clinical systems still persist. In fact, ML models often fail to capture and reconcile the real and complex dynamics of tumors due to reductionist approaches to the problem. Moreover, due to the high heterogeneity of complex omics information, available data can not be enough for classical ML models to properly solve their structure at a granular level \citep{Cordier2021BiologyAM}. Facilitated by the technological innovation and rapid reduction in sequencing costs, thousands of cancer samples have been profiled so far by pan-cancer studies \citep{Weinstein2013TheCG} revealing high level of heterogeneity both intra- and inter- samples. The inter-samples heterogeneity influences clinical tumor classification, patient prognosis and prediction of response to personalized strategies \citep{CaswellJin2021MolecularHA}. Hence, it is of pivotal importance to take it into account, for an effective application of precision oncology. Breast cancer is of special interest given its high prevalence \citep{Siegel2020CancerS2} and its complex landscape of somatic alterations which gave rise to multiple and divergent subtype classifications \citep{Curtis2012TheGA,Russnes2017BreastCM}. In this respect, it represents an ideal complex and real use case to investigate the potentiality of QML approaches to disentangle molecular heterogeneity and extract meaningful non trivial patterns and relationships within the considered dataset \citep{Flother2023TheSO}. Indeed, the exploration of QML approaches in biology and medicine is still at an early stage, with most proof of concept studies exploiting QML models for downsized problem in diagnostics and treatment \citep{Flother2023TheSO}. In particular, Quantum Support Vector Machine (QSVM) has been employed as supervised classification method mainly for distinguishing pathological conditions using clinical records \citep{Krunic2021QuantumKF} or biomedical images \citep{Ahalya2022AutomatedSA}. For Cancer case, BC was classified from mammography using Quantum Transfer Learning \citep{Azevedo2022QuantumTL}, from ultrasound images with Quantum Neural Network \citep{Amin2022BreastMC}, and from digitized image features of fine needle aspirate \citep{Moradi2021ClinicalDC, Vashisth2021DesignAA} or Digital Breast Tomosynthesis \citep{RV2024LeveragingQK}  using QSVM. Very few attempts have been made so far with large and diverse datasets as those including omics data. To the best of our knowledge, the only study is by Li\textit{ et al.} \citep{Li2019QuantumPM} who trained a QSVM exploiting annealing-based Ising ML algorithms on human tumour data from the TCGA and compared its performance with conventional ML algorithms in distinguishing cancer versus normal healthy samples or, among BC tumours, in stratifying by hormonal levels (i.e. LumA vs LumB, ERpos vs ERneg), and in reducing sample complexity. Among the possible quantum algorithm that promise to be disruptive for treating complex data, Quantum Kernel (QK) are regarded as highly efficient tools for identifying non-linear data relationships, by measuring the similarity in a given higher dimensional representation space, which is native when encoding input data in a quantum state{\citep{Schuld2018QuantumML}}. In this study, we make use of QK methods with complex multi-omics tumoral data, aiming to test their potential to detect a more granular structure of the tumoral samples and seek new subgroups, approaching the problem in an unsupervised way. Since the embedding of classical data in a proper way is a highly non-trivial task \citep{Incudini2022AutomaticAE}, we investigated different QK setups that modulate the expressivity of the embedding itself. We also tested various sampling of the BC multi-omics dataset, runnnig the experiments on both simulated and real quantum hardware to assess the consistency of our findings and, finally, we compared them with the classical kernel counterpart. The results show that the quantum approach foster a clusterization that better fit the data with higher number of cluster. Moreover, we noticed that performances trend remains consistent across sample sizes with higher coherence for higher scoring quantum embeddings, suggesting that, for those cases, the patterns extracted can be learnt with fewer data points. We furthermore demonstrated that quantum noise, which in principle hinders the quantum computation, has little or no effect on the computational pipeline for some configurations, thus showing that, in principle, the approach shown here can be implemented on NISQ devices for practical real-world applications.

\section{Background}\label{background}

\subsection{Kernel theory}
In kernel theory, given a set of training inputs $x$ belonging to the input space $X \in R^n$, to recognise patterns a similarity measure  called kernel $k$  \citep{Williams2001LearningWK} is defined as a function
\begin{align}
k: X \times X \rightarrow \mathbb{R}
\label{eq:kernel_func}
\end{align}
\\
Let $\mathcal{H}$ be a Hilbert space, called the feature space, and $x$  a sample from an input set $X$. We can define a feature map $\Phi$\\
\begin{equation}
    \Phi : X \rightarrow \mathcal{H}
    \label{eq:featuremap}
\end{equation}

as a map from inputs to vectors in the Hilbert space. The inner product of two inputs embedded in the feature space define a kernel $k$:
\begin{equation}
    k(x_{i}, x_{j}) := \langle \Phi(x_{i}), \Phi(x_{j}) \rangle
    \label{eq:kerne_inner_prod}
\end{equation}

For a set of $N$ input points $\in X$ a  positively definite matrix called Kernel matrix (or Gram matrix) $K \in R^{N \times N}$ is defined using the kernel function $k$
\begin{equation}
    K := [k(x_{i},x_{j})]_{ij}
    \label{eq:K_matrix}
\end{equation}
for all $(x_{i}, x_{j}) \in X$. The power of kernel methods is that the feature space is usually of much higher dimension than the original space, sometimes even of infinite dimension, where however  the inner product can still be computed through efficient manipulation of input data. An effective and commonly use example is the Radial Basis Functions (RBFs) kernel that projects the input in an infinite dimensional feature space by  evaluating the kernel matrix as follows:
\begin{equation}
    K(x_{i},x_{j}) = e^{- \gamma||x_{i}-x_{j}||_{2}}
    \label{eq:rbf}
\end{equation}
where $||x_{i}-x_{j}||_{2}$ is the Euclidean distance between the two input vectors and $\sigma$ being the standard deviation of the Gaussian distribution. 
QK are kernels where the kernel  is evaluated through inner product of quantum states, since the process of encoding inputs in a quantum state is interpreted as a feature map that embed data to the quantum Hilbert space \citep{Schuld2018QuantumML}. In QK the input data $x \in R^n$ are encoded, via an encoding function  $\phi(x)$, in a quantum state. Formally this encoding process is realized through the application of a unitary transformation $U_{\phi}(x)$   acting on  $n$ qubits  in the initial ground state $|0 \rangle^{\otimes n}$\citep{Mengoni2019KernelMI} :   \\
\begin{equation}
\phi : \textbf{x} \longrightarrow |\phi(x) \rangle= U_{\phi}(x)|0 \rangle^{\otimes n} 
\
\end{equation}
 The kernel matrix is then computed as the pairwise inner product of each quantum state in the dataset:
\begin{equation}
\begin{split}
 K(x_i,x_j) & =|\langle \phi(x_i)||\phi(x_j)\rangle|^2\\
           & =  \langle 0|^{\otimes n}U_{\phi(X_i)}^{\dag}U_{\phi(x_j)}|0 \rangle^{\otimes n}
 \end{split}
 \label{eq:quantum_k}
\end{equation}
Where the QK matrix entries are highly influenced by the choice of the unitary transformation $U_\phi$ \citep{Suzuki2019AnalysisAS}.
\subsection{Quantum Embedding of classical data}
 Classical data can be encoded into the state $ |\phi(x) \rangle$ in several ways and the design of  $U_{\phi}(x)$ should be chosen to maximally exploit the quantum setting, yielding to  classically intractable kernels that are hard to simulate but easy to implement on a quantum computer \citep{Incudini2024TowardUQ}. For this reasons we adopted same  IPQ-style encoding as in \citep{Havlek2018SupervisedLW}. More formally the unitary ansaz we employed is generally described as follows:
 \begin{equation}
 U_\phi(x)=U_z(x) H^{\otimes n}
 \end{equation}
 Where there is a layer of Hadamar gates H acting in parallel on the $n$ qubits, followed by a unitary $U_Z$ that, for entangled embeddings in\citep{Havlek2018SupervisedLW}, takes the form:
\begin{equation}
    U_Z(x)=\prod_{[i,j] \in S} R_{Z_iZ_j}(x_ix_j)\bigotimes^n_{k=1} R_Z(x_k)
\end{equation}
Where $R_{ZZ}$ is a two qubits rotation that makes them entangled and $S$ is the set of qubits to be entangled. Nevertheless, this choice does not guarantee alone an advantage over classical embedding as it is deeply influenced by the data characteristics and tasks \citep{Incudini2022AutomaticAE}. In this respect, we employed the geometric difference($g_{C|Q}$), introduced by Huang et al. \citep{Huang2020PowerOD}, to  evaluate if the classical and quantum kernels implemented differ. This is an anti-symmetric difference between quantum and classical kernel functions that allows to perform a prescreening, independent of the labels,  to measure the separation between the two models for the particular dataset used.The specifics of the quantum embeddings chosen for our experiments are described in the next "Experimental set up" section.
\subsection{Expressivity regularization}
The inherent  high-dimensionality of the Hilbert space used for projecting the input data can hinder the model performance. In particular for QK it has been shown that, as a result of highly expressive problem agnostic encoding, the non-diagonal entries distribution often skews around the mean \citep{Thanasilp2022ExponentialCA}, making the points in the Quantum Hilbert space nearly indistinguishable. This phenomenon has a close parallelism to the curse of dimensionality of classical ML. It has been shown \citep{Shaydulin2021ImportanceOK,Canatar2022BandwidthEG} that acting on the  parametric rotational gate angles can regularise the model expressivity. This strategy is implemented with a tunable hyperparameter called \textit{bandwidth}, which is a constant within the range $[0,1]$. This parameter effectively scales the input, confining the embeddings to a more restricted region of the quantum state space as in: 
\begin{equation}
|\phi(x) \rangle= U_{\phi}(cx)|0 \rangle^{\otimes n} 
\label{bandwidth}
\end{equation}
were $c$ is the bandwidth parameter. 
\section{Experimental set-up}
\label{exp-set-up}
\subsection{Breast Cancer Multi-omics dataset}
The Molecular Taxonomy of Breast Cancer International Consortium (METABRIC) dataset \citep{Curtis2012TheGA} includes the multi-omics profiles of 1980 primary biopsy samples of BC patients and represents the first large effort for studying the BC molecular heterogeneity. METABRIC dataset offers valuable insights, as it comprehensively encompasses a variety of multi-omics profiles which include gene expression and copy number variants. These two omics count respectively 20603 and 22544 features and have highly distinctive properties: gene expression levels are positive continuous variables indicating the abundance of mRNA molecules, whereas CNVs are categorical features representing somatic alterations which involve the changes in the number of copies of a particular DNA segment. In addition, intra-modal and inter-modal independence is not satisfied, and both modalities are subject to the intrinsic noise of the experimental procedure employed to extract these molecular features \citep{Bersanelli2016MethodsFT}. \\
The integration of the genomic and transcriptomic profiles revealed 10 subtypes, termed integrative clusters (IntClust), characterised by distinct genomic drivers \citep{Curtis2012TheGA}. These subtypes hold significant clinical relevance, as they exhibit varying patterns of survival \citep{Curtis2012TheGA}. It is important to note that multiple interpretations of the same dataset have been proposed so far based on alternative intrinsic subtypes (e.g.luminal A, luminal B, HER2-enriched, basal-like or triple-negative, and normal-like), and therefore, the IntClust categorisation should not be viewed as overly restrictive \citep{CaswellJin2021MolecularHA}.\\
Due to the high dimensionality of the dataset, we performed a reduction step formally projecting our input data in a lower dimensional space (i.e. 4 dim). To avoid flattening of meaningful signals through this stage, we used Uniform Manifold Approximation and Projection (UMAP) \citep{McInnes2018UMAPUM} as a reduction strategy since it exploits Topological Data Analysis (TDA) ensuring the topology of the underlying data  is captured and maintained in the reduced space. The algorithm effectively uses a set of hyperparameters to balanced local and global structure; for our case we implemented the transformation with the default parameters.
\subsection{Data encoding: quantum and classical}
\begin{figure*}
    \centering
    \includegraphics[width=1 \textwidth]{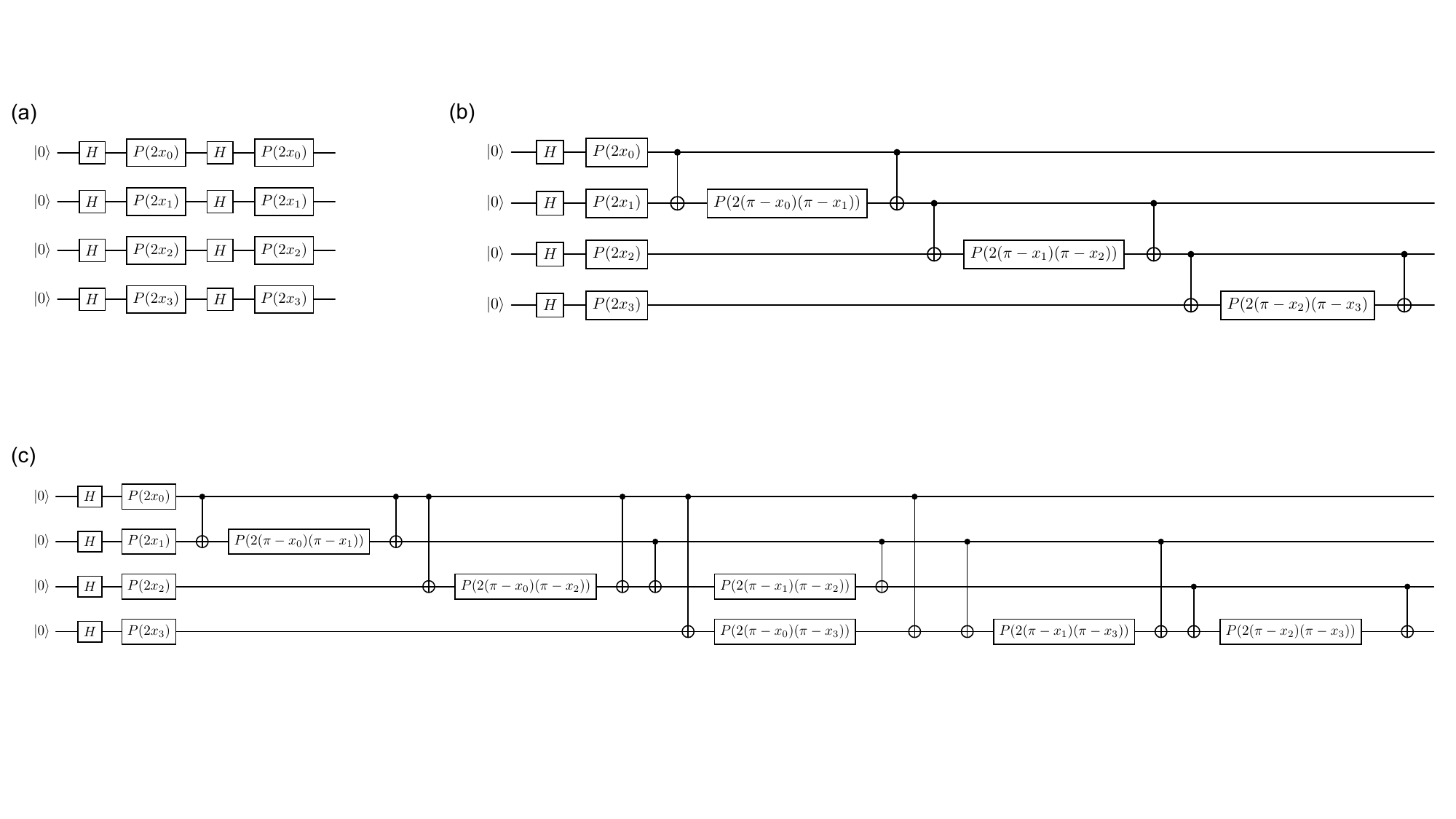}
    \caption{Graphic high-level circuit representation of the four qubits feature maps used with different level of entanglement: (a) Z feature map with no entanglement, (b) ZZ feature map with linear and (c) full entanglement. }
    \label{fig:embedding}
\end{figure*}
The choice of a suitable quantum embedding, in case of complex real data, is not a straightforward task \citep{Cerezo2022ChallengesAO}. Nevertheless, it is possible to explore, as set out in this work, a sub set of configurations of the enconding scheme, in particular by varying the entanglement map of the unitary ansatz. In this way, a trade-off between the expressivity of the kernel and the task-related performance, can be assessed. We specifically explored three configuration of the IPQ-style encoding scheme, making use of unitary templates  with different levels of entanglement of the qubits. The most simple configuration employed is the Z feature map without any entanglement, utilising a sequence of Hadamard and rotational $P$ gates, a single qubit rotation around the $Z$ axis, implemented in our case for a 4-dim classical input $x$ as shown in Fig. \ref{fig:embedding}(a). We then implemented the ZZ feature map with both linear and full entanglement: in the linear entanglement, each qubit is entangled with its neighbouring qubits, while in the full entanglement, each qubit is entangled with every other qubit. These feature map resulted  respectively  in the embedding shown in Fig.\ref{fig:embedding}(b) and Fig. \ref{fig:embedding}(c). Furthermore, we regularised the expressivity of the quantum embeddings tuning the angle range $\beta$. For each feature map, we tested five different scaling of the angle range so that the input data $x \in [0,\beta]$, where $\beta \in \{\frac{\pi}{8},\frac{\pi}{4},\frac{\pi}{2},\pi,2\pi\}$.
To compare the quantum embeddings with a classical kernel  we compute the RBF kernel, expressed by eq. \ref{eq:rbf},  using the sklearn implementation with default parameter $\sigma=\sqrt{\frac{n}{2}}$  where $n$ is the number of feature \citep{scikit-learn}. We selected the RBF kernel  due to its superior expressiveness compared to other classical kernels, enabling it to effectively capture complex, nonlinear relationships in the data \citep{Scholkopf2001LearningWK} .

\subsection{Clustering evaluation}
Clustering is a widely used approach in complex biological data analysis to group similar samples based on molecular descriptors, and  to uncover relationships between cases. Here we choose to adopt Spectral Clustering method to stratify BC samples as it relies on the similarity matrix of the input data like the one we obtain from the kernel matrix (eq.\ref{eq:quantum_k}). 
Diverse internal validation metrics exist to evaluate clustering robustness in an unsupervised case. Each of them is differently influenced by data characteristics (i.e. noise, density, subclusters, skewed distributions) \citep{Liu2010UnderstandingOI}, and has to be interpreted by considering which entities are depicted by the data, as for biological dataset \citep{Ronan2016AvoidingCP}. In our case, with access only to the kernel matrix, we opted for the Silhouette Coefficient (SC) \citep{Rousseeuw1987SilhouettesAG} which relies solely on the pairwise distance between points, and is defined as:
\begin{equation}
SC=\frac{1}{N} \sum^{N}_{i}s(i)
\label{silhouette}
\end{equation}
given a data set of $N$ points and $s(i)$ being the silhouette value associated to the $i$-th point which is defined as follows:
\begin{equation}
    s(i)=\frac{b(i)-a(i)}{\max_i(a(i),b(i))}
    \label{s_index}
\end{equation}
In which $a(i)$ and $b(i)$ are respectively the average distance of  the $i$-th point to its own cluster, and the average distance to its second closest cluster. The value of the both the SC and the $s(i)$ ranges in the interval $\{-1,1\}$ where a high value indicates that the object is well matched to its own cluster and weakly matched to neighbouring clusters. In addition, we employed Adjusted Mutual Information (AMI) as a measure of agreement between two sets of labelled data \citep{Nguyen2009InformationTM}. Given two partitions $U$ and $V$ of the same set of $N$ points, the AMI between them is calculated as:
\begin{equation}
AMI(U,V)= \frac{MI(U,V)-E\{MI(U,V)\}}{\max\{H(U),H(V)\}-E\{MI(U,V)\}}
\label{eq:AMI}
\end{equation}
where H is the Shannon entropy over R clusters and MI is the mutual information between the two partitions, respectively defined as:
\begin{equation}
H(U)=-\sum_{i=1}^{R} P_U(i)\log P_U(i)
    \label{eq:shannon entropy}
\end{equation}

\begin{equation}
MI(U,V)=\sum_{i=1}^{|U|}\sum_{j=1}^{|V|} \frac{|U_i \cap V_j|}{N}\log \frac{N|U_i \cap V_j|}{|U_i||V_i|}
    \label{eq:MI}
\end{equation}

\section{Results} 
\subsection{Exact simulation}
\begin{figure}[h!]
     \centering
     \includegraphics[width=0.49 \textwidth]{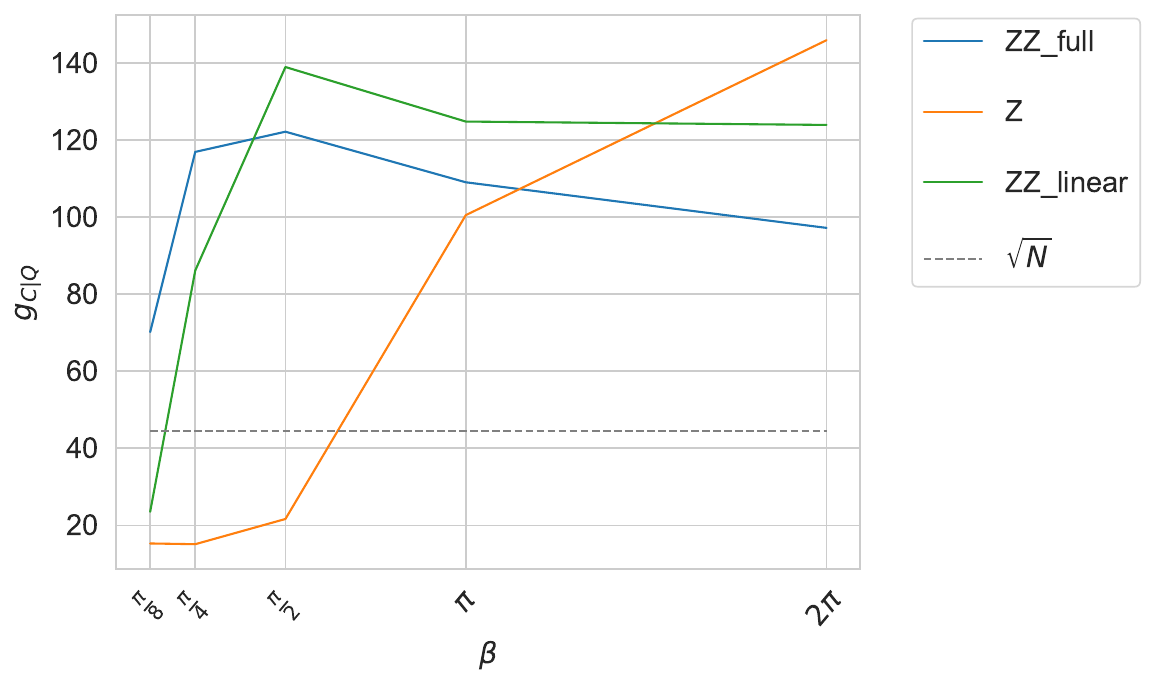}
     \caption{Geometric difference  $g_{C|Q}$ between classical and quantum kernels of the full dataset (N=1980) plotted against the angle range $\beta$  and colored by feature map. The dashed gray line denotes the threshold for a significant difference between the kernels.}
     \label{fig:gdifference}
 \end{figure}
\begin{figure*}[t]
     \centering
     \includegraphics[width=\textwidth]{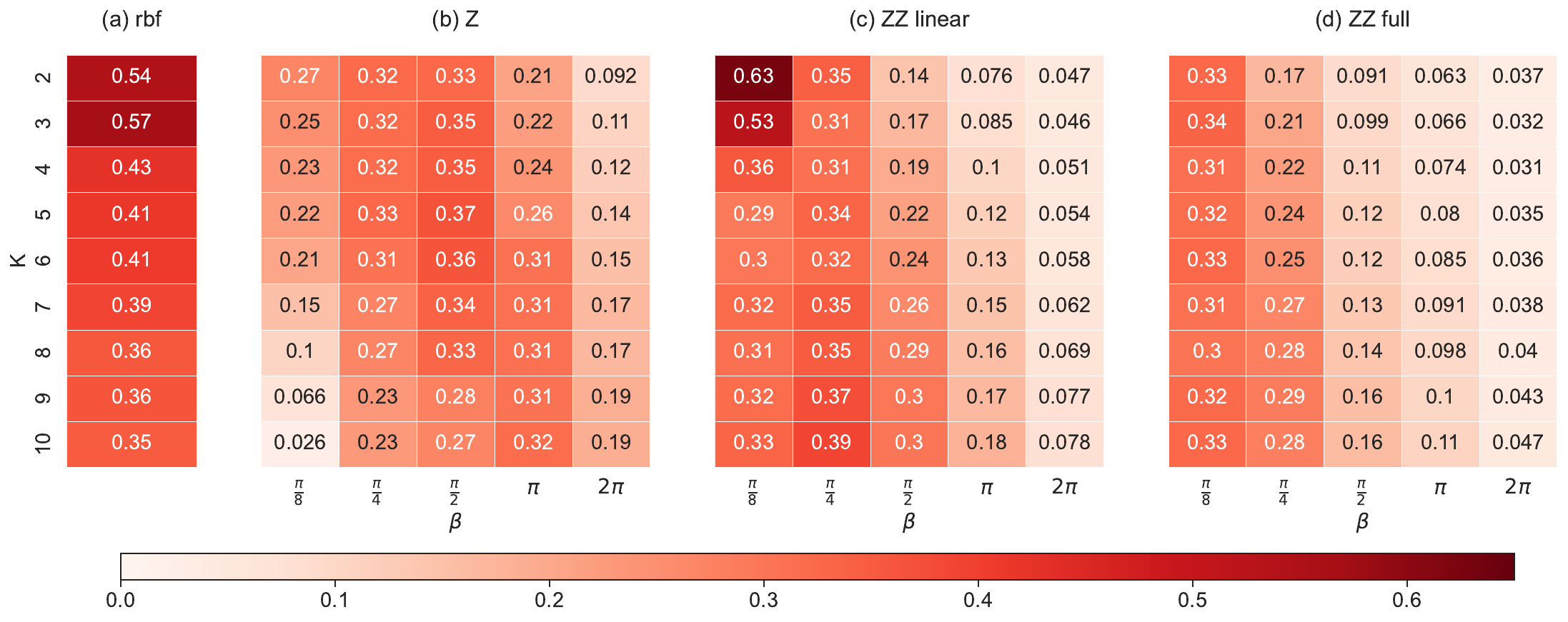}
     \caption{Heatmaps with the SC at varying numbers of clusters $K$ for each feature map configuration: (a) classical RBF; (b) Z encoding; (c) ZZ with linear entanglement; (d) ZZ with full entanglement. For quantum feature maps, SC values are reported at varying of the maximum angle range $\beta$}
     \label{fig:heatmap_SS}
 \end{figure*}
 
We analyzed the performance of the chosen quantum embeddings in an ideal noiseless setting on the full data set with 1980 samples and compared the results with a classical RBF embedding. In this phase we computed the QKs on a locally simulated quantum computer with 4 qubits, using the \textit{qasm simulator} backend from Qiskit , which is designed to simulate the behavior of an ideal quantum circuit. For each feature map the QKs have been evaluated for input values within the range $[0,\beta]$, with $\beta \in \{\frac{\pi}{8},\frac{\pi}{4},\frac{\pi}{2},\pi,2\pi\}$. We investigated the difference between the RBF and each quantum kernel function by computing the $g_{C|Q}$, using the implementation provided by \citep{dimarcantonio2023quask}. The  $g_{C|Q}$  is used as a prescreening of quantum embedding, in particular a large $g_{C|Q}$ compared to the threshold $\sqrt{N}$ indicates a significant deviation between the performances of the two kernels \citep{Huang2020PowerOD}.
The results in Fig. \ref{fig:gdifference} demonstrate that the selected quantum embeddings exceed this threshold, highlighting a significant deviation from the classical counterpart. This behavior correlates with the expressivity of the encoding: the entangled ZZ linear and ZZ full feature maps show $\beta \geq \frac{\pi}{4}$  have a large $g_{C|Q}$  whereas the unentangled Z feature map exceeds the threshold only for less regularized embedding with$\beta \geq \pi $ .  Considering  no assumptions were made regarding the partition of our data,  for the clustering we optimized the number of clusters (K) in a range from 2 to 10, where $K=10$ refer to the IntClust subtypes \citep{Curtis2012TheGA}. Specifically, we measured the SC, as defined in eq \ref{silhouette}, for each feature map (i.e. quantum and RBF) .The results of the SC for the exact simulations are shown in Fig. \ref{fig:heatmap_SS}.
Overall, it can be noticed that ZZ-linear map outperforms the classical RBF model for $K=2$, the trivial case, and for finer grouping with $K\geq 9$,  with $\beta=\frac{\pi}{8}$ and  with $\beta=\frac{\pi}{4}$  respectively. However, the RBF kernel shows higher SCs for $3\leq K\leq 8$. This seemingly irregular behaviour can be explained taking into consideration that the SC might be affected by the presence of macro and micro structures within the data, advantaging clustering with lower $K$ \citep{Liu2010UnderstandingOI}. 
The SC of ZZ full also show an improvement as $K$ increases for all $\beta$ values,  although significantly lower than ZZ linear case, suggesting that a more granular stratification better suits the data structure. The Z feature map, on the other hand, favours clustering with lower $K$ for $\beta \leq \frac{\pi}{2}$ being $K=5$ the best SC value. The trend then switches for $\beta > \frac{\pi}{2}$ for which higher values of $K$ yield a better  SC. In addition, the experimental results demonstrate that the quality of the clustering, entailed by the SC, depends on the complexity of the encoding, i.e. on the entanglement structure of the feature map, and further by the angle $\beta$, which provides a regularisation of the expressivity. In fact, the entangled $ZZ$ feature maps strongly benefit from regularisation,  as they show a subsequent decrease of the score with the increase of the angle range $\beta$. Whereas the Z feature map provides its bests results with some level of angle tuning for $\frac{\pi}{4}\leq \beta \leq \pi$ and has its pick for $\beta = \frac{\pi}{2}$. 
\subsubsection{Sample complexity analysis}
\begin{figure*}[t]
\includegraphics[width=\textwidth]{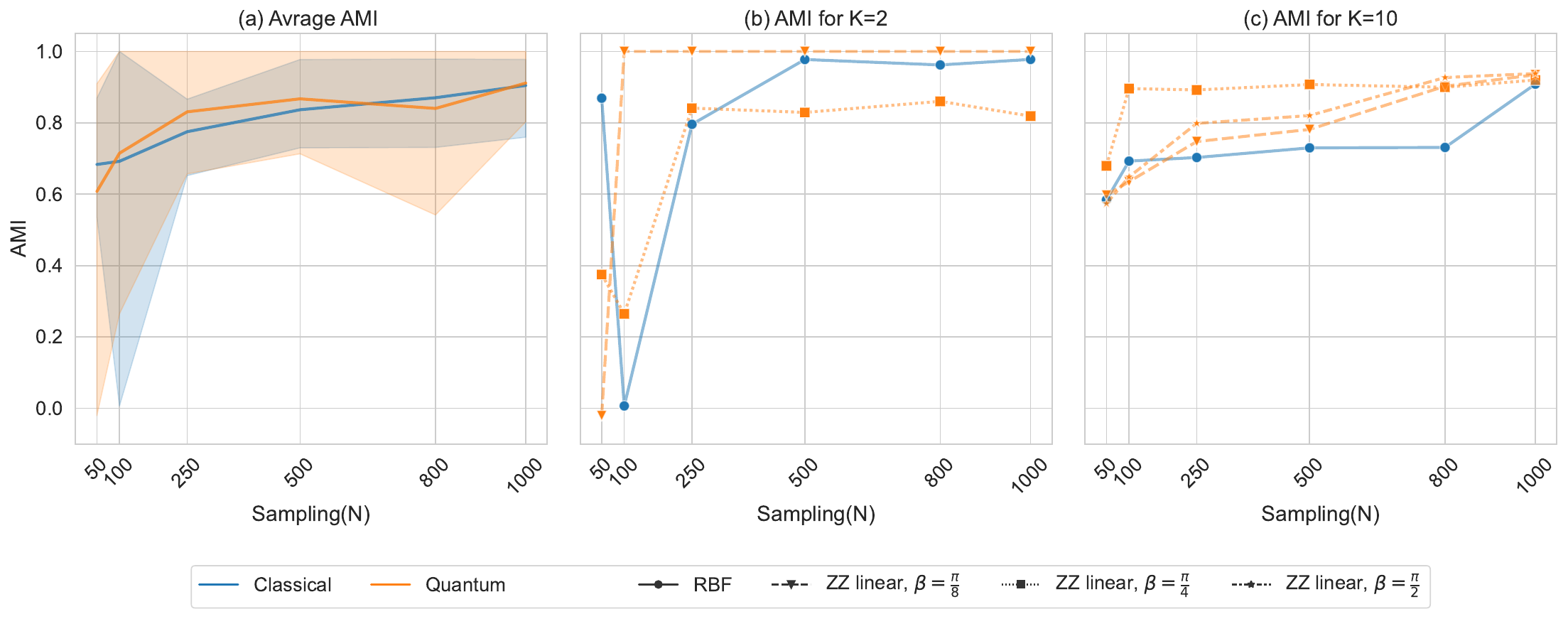}
\caption{Adjusted Mutual Information (AMI) score plotted against the sampling of the data. Results are relative to classical  and quantum stratifications, respectively reported in blue and orange, for configuration with $SC \geq 0.3$. (a) Mean AMI of classical and quantum kernels. (b) AMI for configuration with $K=2$ and (c) $K=10$ }.
\label{fig:AMI_sample_c}
\end{figure*}
To further assess the sample complexity of quantum models for our data, we focused on the best performing ZZ linear feature map. We selected the highest scoring  embedding configuations by setting a threshold of $SC \geq 0.3$, and inspected how samples are grouped among clusters when the sample size varies from $1980$, the full set, to a subset of $N$ samples with $N \in \{50,100,250,500, 800, 1000\}$. We estimated clustering coherence of  the reduced sets by calculating the AMI with respect to the full data for both classical RBF and quantum feature maps. The results are reported in Fig. \ref{fig:AMI_sample_c}. On average, the quantum derived clustering  reach a plateau with an AMI $\geq 0.8$  at  $N=250$ (Fig. \ref{fig:AMI_sample_c} (a)) , slightly before the classical one.  For the trivial case $K=2$ , which presents the highest SCs for both the classical and the quantum case, the $ZZ$  linear map with $\beta=\frac{\pi}{8}$ reaches perfect agreement (AMI = 1) at $N=100$  whereas the RBF kernel needs  $N=500$ samples (Fig. \ref{fig:AMI_sample_c} (b)).  Similar outcomes can be observed with a more granular grouping of the samples with still high SC (i.e.   $K=10$), in which the best configurations of quantum clustering achieve higher AMI compared to the classical one for each sample subset with $N < 1000$. Moreover, the $ZZ$ linear with $\beta=\frac{\pi}{4}$, with the highest $SC=0.39$,  shows a high coherence level  with $AMI=0.9$ even with very few samples (i.e. $N=100$)  (Fig. \ref{fig:AMI_sample_c} (c)). This behaviour suggest that, for those quantum embeddings  that achieve higher scores, the captured structure can be learnt with fewer data points.
\subsection{Execution on QPU}
\begin{figure}[h!]
    \centering
    \includegraphics[width=0.49 \textwidth]{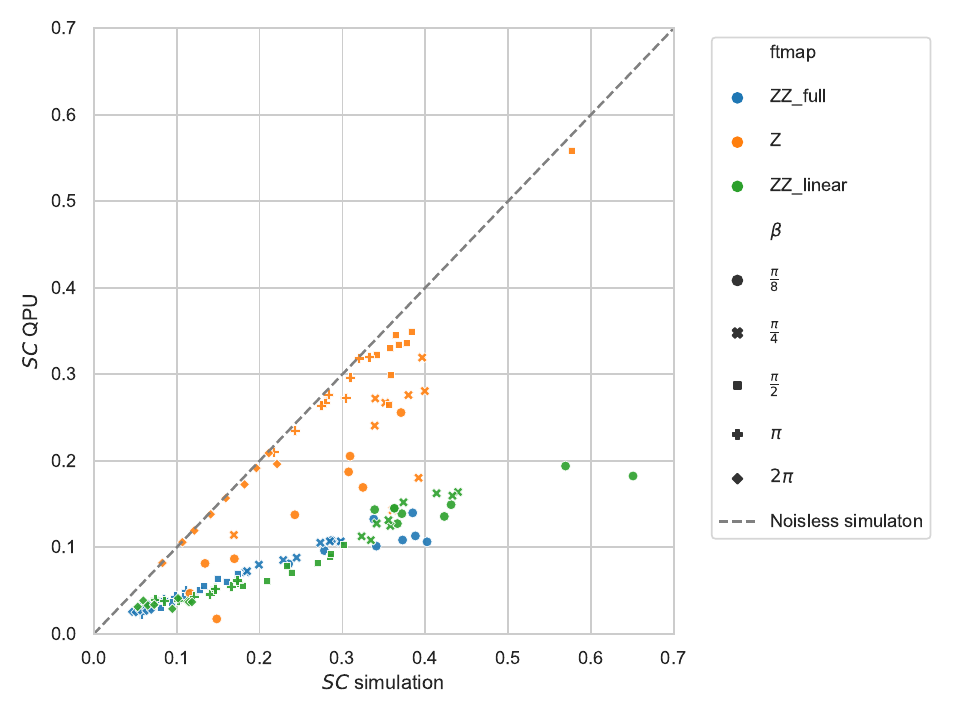}
    \caption{ Comparison of the SC values obtained on the QPU with the noiseless simulation. SC values from QPU plotted against SC values from simulated runs. Point are coloured by feature map and shaped by angle range $\beta$. The dotted grey line denotes perfect consistency with the ideal simulation. }
    \label{fig:SS_coherence}
\end{figure}

Moving forward from the ideal  simulation, we computed all QK on a real noisy NISQ device, to assess the reliability of numerical simulation in light of quantum noise, and verify how  clusters might vary between different embeddings. The current limitations on available systems hinder the ability to run the experiments on the whole dataset, thus we consider, in the following, a reduced sampling of the data with $100$ points. This fact, while is expected to affect the results of the clustering analysis, does not undermine the capacity of addressing the overall effects of the quantum noise on the computational pipeline, as it depends mostly on the hardware specification and on the circuit composition.
\begin{figure}[h!]
    \centering
    \includegraphics[width=0.5\textwidth]{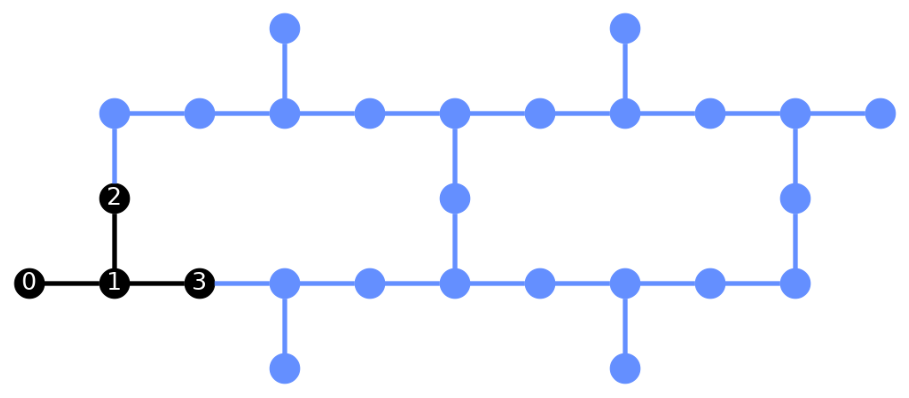}
    \caption{IBMQ\_Mumbai 27 qubits hardware specifics and topology. Qubits selected for circuits implementation are shown in black. }
    \label{fig:ibmq_top}
\end{figure}
Indeed, to run a quantum circuit on a real quantum computer, specifically a gate base quantum hardware, it is necessary to rewrite it from an high-level circuit into an equivalent circuit that  matches the topology of the specific quantum device, and further to optimize the execution for present day noisy quantum systems. This process, called transpilation, can change substantially the outline of the original circuit in terms of  circuit depth, number and type of gates used. Since these aspects are crucial for determining the resilience of a circuit to noise, they need to be taken into account in discussing quantum hardware results. In table \ref{tab:circ_comp} we reported the specific of each transpiled circuit in use.
We implemented our experiments on the ibm quantum system $ibmq\_Mumbai$, fitting the quantum circuits to the hardware topology as shown in Fig. \ref{fig:ibmq_top}: out of the 27 qubits in the device we heuristically selected 4 with higher level of connectivity and that minimise the average readout error rate. 
\begin{table}[h!]
\caption{The table describes the depth and CX gate abundance of the transpiled circuits for each feature map. The numbers are relative to the full circuit which corresponds to $U^{\dag}U$. }
\begin{tabular}{@{}llll@{}}
\toprule
Feature map($U$)& Entanglement type & Circuit depth & CX gate \\ \midrule
Z              & None              & 5             & 0       \\
ZZ             & linear            & 63            & 34      \\
ZZ             & full              & 45            & 24      \\ \bottomrule
\end{tabular}
\label{tab:circ_comp}
\end{table}
For each QK evaluated on  $ibmq\_Mumbai$ we calculated the SC,  here reported in Fig. \ref{fig:SS_coherence}, versus the values obtained in simulation. The Z feature map with $\beta \geq\frac{\pi}{2}$ exihibits very good consistency with the ideal case, whereas the entangled  ZZ linear and ZZ full feature maps deviate more from the ideal scenario. Still the present behaviour can be explained considering the depth and composition of the circuits  transpiled on the QPU: in fact, as reported in table \ref{tab:circ_comp}, entangled feature maps account for a deeper circuit and a higher number of CNOT gates and thus are more susceptible to noise, both single qubit noise and gate-level noise.
Since different stratification can yield to the same SC values, to further assess the clustering coherence we measured the AMI between simulated and real quantum stratification for each combination of the feature map, $K$, and $\beta$.  
\begin{figure*}[t]
    \centering
    \includegraphics[width=1 \textwidth]{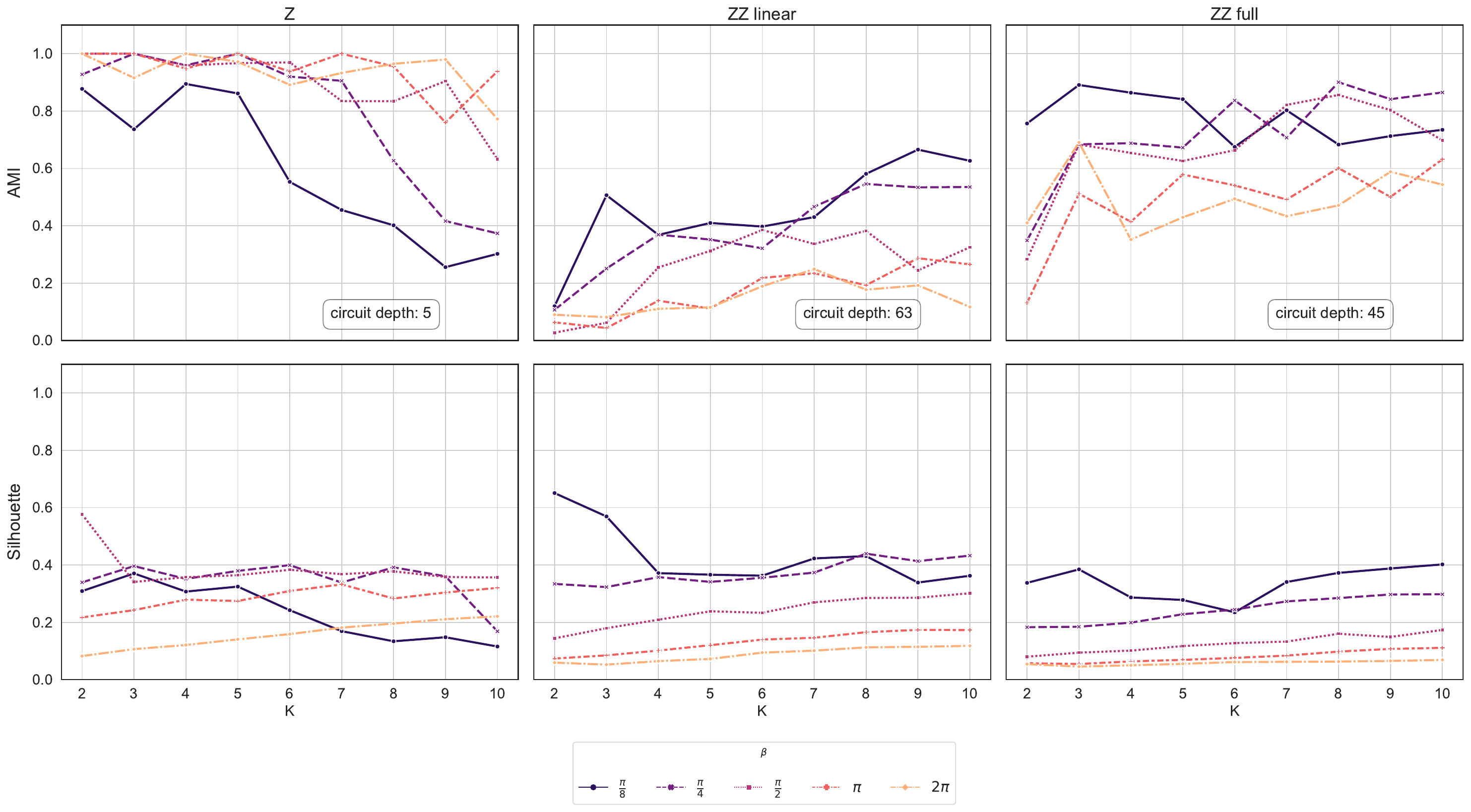}
    \caption{Plots of AMI comparing simulated and real quantum stratification and SC of ideal simulation for 100 samples. The plots in figure are organized in rows by metric (top row AMI and bottom  SC) and in columns by feature map type, from left to rigth: Z, ZZ linear, and ZZ full. The data points are plotted against the number of clusters $K$ and are coloured and shaped by angle range $\beta$}
    \label{fig:AMI}
\end{figure*}

For each feature map we  plotted the AMI results for each $K$ and  $\beta$ values in Fig. \ref{fig:AMI}.  The deepest feature map (circuit depth $= 63$), the ZZ linear, presents the lowest AMI scores, while  the overall best AMI scores distribution are associated with the Z feature map (circuit depth $= 5$), which is in principle the less expressive. The fully entangled feature map (circuit depth $= 45$) shows intermediate values. Thus, the overall trend of the AMI  confirms the previous findings in Fig \ref{fig:SS_coherence}: both the circuits depth and the gate composition emerging from the transpilation process, have a significant impact on the clusters consistency. In addition, using the Z feature map the coherence between simulated and real clusters increases with higher values of $\beta$: in particular, for $\beta > \frac{\pi}{4}$  and $K \leq 7$ , it achieves perfect agreement or almost perfect agreement. In contrast, the entangled feature maps exhibit the opposite trend, where a smaller $\beta$ corresponds to a higher AMI. By crossing these results from real quantum hardware with the SC obtained in the ideal scenario (Fig. \ref{fig:AMI} ), which provides the best possible outcome for these data, it is possible to draw some conclusion about the effect of quantum noise on the clustering. It can be noticed that quantum kernels of entangled feature maps with high $\beta$ give small SCs, thus they do not provide accurate clusters. In these cases the data can be easily partitioned very differently upon perturbations, potentially coming from any noisy source. Hence, when a real quantum hardware is in use, as there is an intrinsic noise source in the computation, the robustness of the clustering provided by the entangled kernel functions, with corresponding low SC values, is strongly hindered by the quantum noise.
\subsection{Limitations}
Our results show that QKs achieve clustering performance comparable to the classical RBF kernel and a significant reduction in sample complexity, still these findings should be considered alongside certain limitations of the designed experimental set-up. Specifically, the outcomes are influenced by the data preprocessing step used to reduce dimensionality but its impact has not been investigated. Future studies should address this in terms of both reduction method and dimension of the projected space. In addition we acknowledge that, while METABRIC has been chosen as a benchmark  for its multi-omics features, high dimensionality and heterogeneity,  our results are dataset dependent and might not uphold to a different multi-omics tumor dataset. In particular, the relationship between the dataset characteristics and the observed advantages of the quantum model, like the reduction in sample complexity, was not quantitatively investigated as it goes beyond the scope of our work, but has emerged as an interesting aspect to study in the future in light of the practical application of QKs. 
\section{Conclusion and Future work}\label{Conclusion}
We have investigated QK models for BC subtyping with a real multi-omics dataset addressing the challenges of problem-agnostic quantum embeddings for the best expressivity task-related trade-off. We explored three IPQ-style encoding schemes with different levels of entanglement of the qubits, comparing them with a classical RBF model. The implemented QKs yield a  $g_{C|Q}$ that consistently exceeds the threshold, thus guarantying a sufficient separation between the quantum and the classical models. The QK obtained with a noiseless simulation on the full dataset ($N=1980$) show that the ZZ linear feature map achieved better clustering results, in terms of SC, compared to the classical RBF kernel for finer stratification of BC samples, with $K \geq 9$. In addition, varying the entanglement level and angle range, we noted how entangled feature map, which are in principle more expressive, strongly benefits  from angle tuning,  confirming that embedding optimisation is needed. Further, we focused on the best data-fitting feature map, the ZZ linear, and demonstrated experimentally that configurations with better clustering results ($SC \geq 0.3$) foster a reduction of sample complexity. Indeed for the best performing cases ($K=2, K=10$) the ZZ linear was able to obtain highly coherent stratification ($AMI \geq 0.9$) with the one obtained for the full dataset using only 100 samples. This finding, already identified by \citep{Li2019QuantumPM} in the supervised setting, is potentially highly valuable for clinical applications, where datasets and sample cohorts are often not sufficiently large and representative due to experimental and availability limitations. While the proposed methodology showed promising results in crucial aspects for possible future applications of quantum models in biomedicine, we have tested here only a limited subset of possible encoding configurations. To further explore the encoding space to enhance performances, our work could be extended by using a trainable kernel approach \citep{Glick2021CovariantQK}. In such case a parametrized QK is iteratively adapted to a dataset. In addition, we acknowledged that, our results are influenced by the preprocessing choices made for our data. Future work could investigate the impact of dimensionality reduction on performance, exploring both different reduction strategies and varying the dimensionality of the projected space (e.g. extending beyond the 4-dimensional space used in this study). Finally, the execution on a NISQ system, the \textit{ibmq\_Mumbai}, showed that quantum noise has little to no effect on less expressive configurations, like the Z feature map that entails shallower circuits, while it hinders the results of entangled feature maps. This behaviour suggests that a trade-off between ideal performances and hardware resilience should be considered in quantum kernel selection. Nevertheless is important to note that NISQ devices now available are rapidly evolving and have reached higher accuracy level, compared to the QPU used in this work \citep{Kim2023EvidenceFT}. In addition, a major effort has been made also in advancing noise mitigation strategies\citep{Cai2022QuantumEM}, thus we expect that, in the foreseeable future, quantum protocols, like the one set out in this work, will achieve sufficient reliability and scalability to be implemented for real applications.\\
The clustering of  tumour samples using multi-omics data is crucial to identify clinically relevant subgroups, still the reconstruction of non linear relationship between tumoral molecular descriptors is made hard by classical model expressive capacity and  limited data. Although QML and QK could offer an alternative view,  we faced some major challenges in building proper quantum representation for biomedical systems, where formulating a problem-inspired  embedding (eg. \citep{Larocca2022GroupInvariantQM, Liu2020ARA,Meyer2022ExploitingSI} ) is not possible. Nevertheless, similar considerations apply to classical ML, and given the early stage of quantum approaches, the heuristic exploration of  classically complex problems is necessary to identify tasks that could be tackled with current NISQ devices. In this spirit, we would like to further explore the use of QKs for challenging biomedical problems that could benefit from a quantum modeling of the system, such as peptide binding prediction, for which an embedding could be more easily formulated based on \textit{a-priori} knowledge of the system.

\section*{Acknowledgement}
We acknowledge the CINECA award under the ISCRA initiative, for the availability of high performance computing resources and support.
\newline
\section*{Declarations}
\subsection*{Funding}
This research was financially supported by the Consortium GARR fellowship "Orio Carlini" (to VR), the European Union - Next Generation EU, in the context of PNRR, Investment 1.5 Ecosystems of Innovation, Project Tuscany Health Ecosystem (THE), ECS00000017, Spoke 3. CUP: B83C22003930001 and IIT-CNR Consolidator grant (to VR and RDA), the PNR MUR project PE0000013-FAIRn (to GB). 
\subsection*{Conflict of interest}
The authors have no competing interests to declare that are relevant to the content of this article.
\subsection*{Data and code availability }
METABRIC dataset, including gene expression andcopy number data, for the 1980 patients is  freely available at cBioPortal repository, at the following link:\url{ www.cbioportal.org/study/summary?id=brca_metabric}\\
The data, code, and experimental results associated with this project is available on GitHub at the following link: {\href{https://github.com/ctglab/QKomics}{https://github.com/ctglab/QKomics}}
\subsection*{Authors contribution}
VR conceived and conducted the study, performed computation and analyses, interpreted the results, and drafted the manuscript; EGC contributed in developing simulation models and first conceptualization; GB contributed to the conceptualization, verified analytical methods, and drafted the manuscript; RDA conceived and supervised the study, interpreted the results, and drafted the manuscript. All authors reviewed and approved the final manuscript.
\bibliography{bibliography}

\end{document}